\documentclass[a4,notitlepage,12pt]{jedm}
\usepackage[table]{xcolor}
\usepackage{url}
\usepackage{graphicx}
\usepackage{amsmath}
\usepackage{booktabs}
\usepackage{hyperref}
\usepackage{multirow}
\usepackage{geometry}
\usepackage{float}
\usepackage{threeparttable}
\usepackage{booktabs}
\usepackage{subcaption}

\begin{document}

\title{\fontsize{23}{14}\selectfont E-TRIALS: Empowering Data-Driven Decisions to Enhance Computer-Based Learning Platforms}
\date{}
\newcommand{\authorFixedWidth}[1]{\parbox[t]{.25\textwidth}{\raggedright#1 \raisebox{0pt}[0pt][6pt]{}}}

\author{
  \begin{tabular}{@{}ccc@{}}
    \parbox[t]{.33\textwidth}{
      \authorFixedWidth{{\large Abubakir Siedahmed}\\Worcester Polytechnic Institute\\Worcester, MA, USA\\asiedahmed@wpi.edu}
    } &
    \parbox[t]{.33\textwidth}{
      \authorFixedWidth{{\large Yanping Pei}\\Worcester Polytechnic Institute\\Worcester, MA, USA\\ypei@wpi.edu}
    } &
    \parbox[t]{.33\textwidth}{
      \authorFixedWidth{{\large Adam Sales}\\Worcester Polytechnic Institute\\Worcester, MA, USA\\asales@wpi.edu}
    } \\
    \parbox[t]{.33\textwidth}{
      \authorFixedWidth{{\large Neil T. Heffernan}\\Worcester Polytechnic Institute\\Worcester, MA, USA\\nth@wpi.edu}
    } &
    \parbox[t]{.33\textwidth}{
      \authorFixedWidth{{\large Johann A. Gagnon-Bartsch}\\University of Michigan\\Ann Arbor, MI, USA\\johanngb@umich.edu}
    } &
    \parbox[t]{.33\textwidth}{
      \authorFixedWidth{{\large Di Zhang}\\University of Utah\\Salt Lake City, Utah, USA\\Di.Zhang@utah.edu}
    }\\
    \parbox[t]{.33\textwidth}{
      \authorFixedWidth{{\large Brendan A. Schuetze}\\University of Utah\\Salt Lake City, UT, USA\\brendan.schuetze@utah.edu}
    } &
    \parbox[t]{.33\textwidth}{
      \authorFixedWidth{{\large Allison Zengilowski}\\Lehigh University\\Bethlehem, PA, USA\\alz324@lehigh.edu}
    } &
    \parbox[t]{.33\textwidth}{
      \phantom{Dummy}
    }
  \end{tabular}
}

\maketitle

\begin{abstract}
Computer-based learning platforms (CBLPs) have become a common medium in schools, transforming how students learn and interact with educational content. However, researchers still lack adequate tools to address the diverse set of challenges that students face in these environments. In this paper, we introduce \textbf{Ed-Tech Research Infrastructure to Advance Learning Sciences (E-TRIALS)}, a free tool developed by ASSISTments to help researchers conduct randomized controlled trials in the realm of learning sciences. We describe its features, the types of experiments it supports, and how it can address critical research questions. We showcase E-TRIALS' capabilities through two real-world interventions. Finally, we evaluate the efficacy of interventions using three average treatment effect (ATE) estimators. Student's t-test, regression, and Leave-One-Out Potential outcomes (LOOP). The results demonstrate that the unbiased LOOP estimator can achieve greater precision by adjusting for baseline covariates compared to the Student's t test. Our work demonstrates the potential of E-TRIALS to advance research and contribute to the development of more effective, inclusive, and adaptive CBLP. The code used for this work is available at https://osf.io/xp6ch/.
\\
\\
{\parindent0pt
\textbf{Keywords:} Educational Interventions, E-TRIALS, A/B Testing, Learning Analytics, Treatment Effect Estimators
}

\end{abstract}


\section{Introduction}
In 1923, Thomas Edison predicted that motion pictures would revolutionize the education system within 20 years by replacing teachers and textbooks entirely \cite{salisbury1986teachers}. This prediction reflects a recurring pattern in the education system, which is the introduction of new technologies often sparks enthusiasm and bold claims about their potential to transform learning. For instance, in 1961, the Ford Foundation invested over $10$ million US dollars - equivalent to approximately $106$ million US dollars in 2025 - in a program aimed to promote televised instruction. The foundation believed televised instruction could "equal or surpass conventional methods" and make superior teaching resources accessible to more students \cite{FordFoundation_1961}. Some of the funds were even used to release teachers from their daily duties to serve as "television consultants."

This optimism was not unique to television. Similar enthusiasm emerged with the widespread adoption of computers in education. Educators quickly embraced computers for administrative tasks, research, and instructional purposes \cite{billings1988computers}. In the 1960s, the National Science Foundation spearheaded efforts to integrate computers into classrooms aiming to enhance instruction. At the time, only 1\% of secondary schools had access to computers, but by 1975, this figure had risen to 23\% \cite{ComputersInEducation_1997}. Initiatives like the "Computer Literacy Curriculum" were introduced to ensure students and teachers were adequately prepared for this technological shift \cite{Molnar_1975}. 

The growing presence of computers in schools paved the way for intelligent tutoring systems (ITS), which aimed to deliver personalized instruction, provide immediate feedback, and assess student knowledge \cite{billings1988computers}. One of the earliest examples, SOPHIE (SOPHisticated Instructional Environment), was developed in 1975 \cite{Brown1974SophisticatedIE}. SOPHIE simulated electronic equipment and allowed students to practice troubleshooting without the risk of damaging real devices \cite{Brown1974SophisticatedIE}. Leveraging techniques from artificial intelligence, cognitive science, and psychology, SOPHIE was able to understand natural language inputs and respond in a supportive tone within two seconds. Additionally, SOPHIE outperformed human observers in identifying faults in their trouble-shooting reasoning and was more effective than students at diagnosing errors in their reasoning. 

Another notable advancement in ITS was the development of cognitive tutors based on the Adaptive Control of Thought (ACT) theory of learning and problem-solving \cite{anderson1995cognitive}. These tutors were guided by eight principles, such as representing student competence as a set of production rules, providing immediate feedback, and minimizing cognitive load \cite{anderson1995cognitive}. Researchers applied these principles to create tutors for subjects like LISP, geometry, and algebra \cite{anderson1995cognitive}. The LISP tutor helped students complete exercises 30\% faster and achieve 43\% higher scores on a posttest \cite{anderson1995cognitive}. The algebra tutor initially didn't show significant results, but a later version showed significant achievement gains. The geometry tutor helped improve student's grades by more than one letter grade on a test \cite{anderson1995cognitive}. 

Over the years, terms like intelligent tutoring systems (ITS), computer-assisted instruction (CAI), and online learning platforms have often been used interchangeably. While these systems share similarities, such as providing immediate feedback and facilitating learning, they also have distinct features. For instance, ITS are typically knowledge-centered, whereas CAI tends to be interaction-centered \cite{larkin2021computer}. To avoid confusion, this paper uses the term \textbf{Computer-Based Learning Platforms (CBLPs)} to refer to systems that: 1) allow students to interact with simulations or problem-solving tasks, 2) provide personalized or generic feedback, 3) adapt instructional content or assessments to the student's knowledge level, 4) incorporate multimedia elements (e.g., text, audio, graphics, video), and 5) track student progress and generate performance reports for students and teachers. 

The adoption of CBLPs in classrooms has surged in recent years, largely due to the increase of availability of computers to each student \cite{gopika2023awareness}. Before the COVID-19 pandemic, 45\% of public schools reported having a computer for every student \cite{gray2021use}. By the 2022-2023 academic year, this figure had risen to 94\% \cite{IES_SPPSurvey}. Unlike earlier technologies, CBLPs have consistently demonstrated their ability to enhance learning outcomes, particularly when designed to support, rather than replace teachers \cite{anderson2014rules,murphy2020investigating,leelawong2008designing,toyama2011shortcuts}. CBLPs have transformed how students learn, complete assignments, and receive assessments. For example, CBLPs enable students to receive immediate and personalized feedback, while teachers benefit from analytical dashboards that provide insights into student performance. These tools allow educators to quickly identify struggling students and tailor their support accordingly. 

Despite their advantages, developing effective CBLPs is a complex challenge due to the vast array of instructional design choices available. As Koedinger et al., noted in \textit{Science}, there are approximately 300 trillion possible design choices \cite{koedinger_instructional_2013}. Researchers are struggling to identify which instructional choices to include and omit. Instructional choices, such as how often should students receive feedback, and what type of feedback is most effective? Should high-performing students receive delayed feedback, while low-performing students receive immediate feedback? Should instruction rely solely on text, or should it incorporate multimedia elements like videos, graphics, or audio? Are open-ended questions more effective than multiple-choice questions? These are just a few of the many decisions researchers and developers must navigate. 

The overwhelming number of possible design choices birthed the term "the rise of the super experiment" \cite{stamper_rise_2012}. This concept refers to data-driven educational experiments conducted at the internet level to identify which instructional choices are most effective and to whom. To address this need, tools like UpGrade by Carnegie Learning and TerraCotta within the Canvas Learning Management System have been developed to facilitate such experiments \cite{fancsali_closing_2022}. However, TerraCotta is used for experiments within a learning management system like as Canvas or Google Classroom \cite{motz2024terracotta}, and UpGrade is a tool that works by using its API to connect to the CBLP to deliver the educational content and researchers would need to develop their own content \cite{wang2019upgrade}. 

In this paper, we introduce \textbf{Ed-Tech Research Infrastructure to Advance Learning Sciences (E-TRIALS)} a free and open tool developed by ASSISTments. E-TRIALS enables researchers to conduct experiments in any subject area. We describe its features, the types of experiments it supports, and how it can address critical research questions in learning sciences and technologies (LS\&T). To demonstrate the capability of E-TRIALS, we introduce two E-TRIALS experiments and evaluate the effectiveness of the interventions using three Average Treatment Effect (ATE) estimators: (1) Student’s \textit{t}-test ($\hat{\tau}_{\text{t-test}}$), (2) Ordinary Least Squares regression ($\hat{\tau}_{\text{Reg}}$), and (3) Leave-One-Out Potential outcomes estimator (LOOP) ($\hat{\tau}_{\text{LOOP}}$). These three methods are chosen because they represent a progression in complexity and robustness. The \textit{t}-test estimator is the simplest and most intuitive, relying solely on the difference in means between the treatment and control groups. Although RCTs should, in theory, ensure covariate balance on average, real-world RCTs often exhibit residual imbalance. In such cases, covariate adjustment becomes important. Regression is the most commonly used method for this purpose. The LOOP estimator, a more advanced approach, offers an unbiased estimate of the ATE while accounting for covariate differences, even in small samples.

\section{Experimental Environment}
ASSISTments is a web-based and free-to-use CBLP designed to assist teachers, rather than replace them \cite{heffernan_assistments_2014}. The platform allows teachers to assign mathematical problems from open-source curriculum, such as Illustrative Mathematics, or to create their own problem sets. As students work through their assigned assignments, they can request support, which would appear in the form of hints or explanations. ASSISTments is widely used, with over 500,000 students and 30,000 teachers annually \cite{assistments2025}. 

\subsection{E-TRIALS}
The Ed-Tech Research Infrastructure to Advance Learning Sciences (E-TRIALS) is a research tool designed by ASSISTments. It enables researchers to conduct experiments in the realm of learning sciences in any subject without writing a single line of code. Before the wide spread of CBLP and tools like E-TRIALS, researchers relied on manual methods to design and conduct experiments \cite{angrist2003randomized}. E-TRIALS connects researchers with a large and diverse population of students (>500,000) and teachers (>30,000) to conduct experiments. Once a researcher completes designing the experiment in E-TRIALS, they would not need to worry about recruiting participants because ASSISTments does the recruiting for them by assigning its users to the experiment conditions.  
E-TRIALS is one of five digital platforms funded by the Institute of Education Sciences (IES) to develop an infrastructure to support educational research questions \cite{schellinger2024considerations}. Altogether, these platforms - UpGrade, Kinetic, Terracotta, and ASU's Learning@Scale - have helped researchers connect to millions of teachers and learners across K-16 \cite{schellinger2024considerations}.

\subsubsection{Previous Experiments in E-TRIALS}
Since the inception of E-TRIALS, dozens of researchers have used it to conduct RCTs \cite{featured_works}. Walkington et al. conducted a study in E-TRIALS that explored how language skills can impact student performance in algebra and evaluate their effects on accuracy and response time \cite{walkington2019effect}. They manipulated six language features: 1) the number of sentences, 2) pronouns, 3) word concreteness, 4) word hypernymy, 5) sentence consistency, and 6) problem topic. The study had 451 students and 2,825 observations, where each observation represented a student's response to a problem. While the findings showed little evidence that language features significantly impact mathematics word problem-solving performance, the findings did reveal that sentence consistency reduced response time. Another study on algebra explored the effects of feedback for middle school students by testing four conditions across two experiments: 1) students received correct-answer feedback or no feedback, and 2) students received explanation feedback or try-again feedback \cite{fyfe2016providing}. Across a total of 246 students in both experiments, posttest scores revealed that students with low prior knowledge benefited from all feedback types, instead of the no feedback condition. However, all four conditions didn't affect the performance of students with high prior knowledge. Though these two studies demonstrate E-TRIALS' ability to help researchers improve student performance through research-backed interventions, the following study is especially worth mentioning due to the large number of student participants of 2,152 from 5th-12th grades \cite{harrison2020spacing}. Students were randomly assigned to one of four conditions that manipulated the spacing between numbers and terms within mathematical expressions to be either 1) neutral (no extra spaces), 2) congruent with the order of operations (using grouping), 3) incongruent with the order of operations, or 4) a combination of the previous conditions. By using mastery speed (students' total attempt count before correctly answering three consecutive problems) as the outcome measure, the study found that students in the incongruent condition had slower mastery speed than those in the congruent or neutral conditions. These findings suggest that spacing that conflicts with the order of operations presents additional challenges for students when solving mathematical expressions. 

In the above papers, researchers did not have to recruit teachers or students, or even design the problem content. They selected a set of problem set(s) from the ASSISTments library of math content and then incorporated their condition(s) into the problem set(s). Once the study is ready, ASSISTments deploys it and randomizes the students to a condition. E-TRIALS also offers the flexibility for researchers to create their own problem content and recruit participants. A study by \cite{oppenzato2024rebalancing}, noticed that oftentimes addition and multiplication fraction problems in textbooks are often common for some combinations of operand types and operations, while it can be rare for others. In the first iteration of the study, the researcher designed a pretest-intervention-posttest in Qualtrics to find out whether a more balanced distribution of practice problems, especially for rare types of problems could improve student performance. 63 students were randomly assigned to one of two conditions: 1) more common types of problems, or 2) balanced mix common and rare problems. The final data contained 40 students because many were removed for scoring a perfect score on the pretest. Due to a small sample size, the researcher decided to redo the study. The second iteration of the study was conducted in E-TRIALS to ensure a larger sample size and only allow teachers who have a high rate of underperforming students to participate in the study. The researcher developed the content for the pre- and posttest, and the practice problems. Findings from 127 students show that students who received the balanced mix of common and rare problems improved and made fewer independent whole number errors on whole-fraction multiplication items than students who received common types of problems. 

Another study by Zhou et al. demonstrates another advantage E-TRIALS offers researchers \cite{zhou2021comparison}. Zhou et al. investigated how hint and scaffolding tutoring strategies can influence adult learners' interactions within the edX MOOC platform. Using the Learning Tools Interoperability, edX and ASSISTments were integrated to deliver the content via edX and the assignment via ASSISTments. Students in the first condition were able to request hints at any point during a problem. Students in the second condition automatically received a sequence of scaffolding questions if their first attempt on an incorrect problem, and they were required to complete the entire sequence before starting the next problem. Results from the study revealed that scaffolding didn't help adult students. In fact, students with low prior knowledge requested significantly more correct answers and failed to solve later problems.  

In a broader analysis, \cite{prihar2022exploring} aggregated results from 50 experiments that were conducted in E-TRIALS from various researchers. These experiments investigated a wide range of research questions, such as the effectiveness of explanation vs. answer-only feedback, video examples vs. text-based, common wrong answer feedback vs. no feedback, fill-in vs. multiple-choice questions, and motivational vs. non-motivation messages. Collectively, these experiments had 30,408 K-12 students. The primary goal of this meta-analysis was to identify consistent effective patterns across experiments, such as personalization and effective tutoring methods within CBLPs. 

As demonstrated by the interventions discussed above, E-TRIALS has enabled researchers to explore researcher questions with minimal effort. Its flexibility coupled with the types of data the platform collects has allowed researchers to offer researched-backed design choices for CBLP, and in doing so, helped learners perform better academically. E-TRIALS opens new avenues for addressing novel research questions and generating data-driven insights. Its ability to handle large sample sizes, diverse interventions, and complex experimental designs makes it an invaluable tool for advancing research. 

\subsubsection{Developing an Experiment}
\begin{figure*}[htbp]
        \centering
        \includegraphics[width=0.9\linewidth]{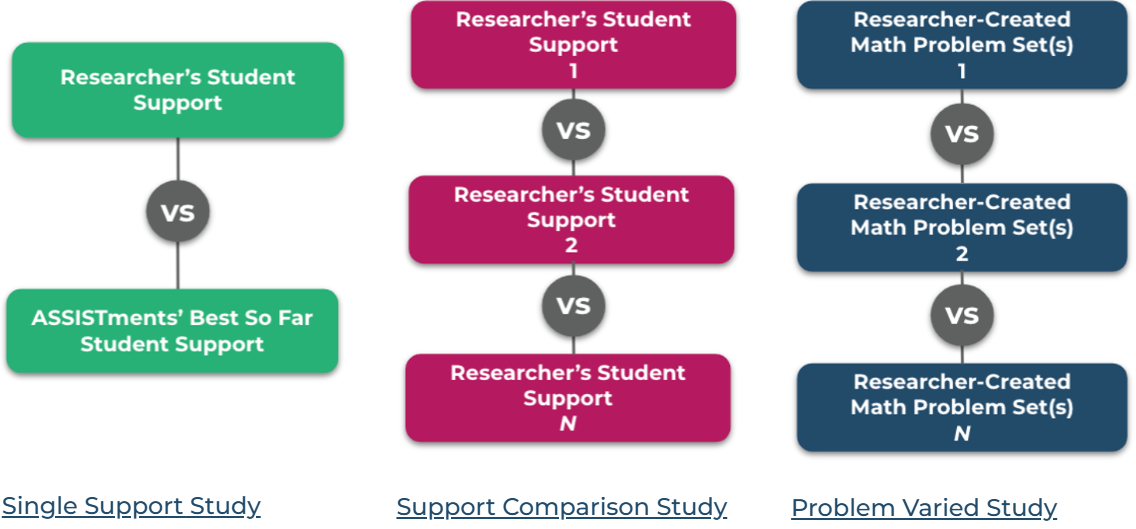}
        \caption{The 3 Types of Studies in E-TRIALS}
        \label{fig:Figure 1} 
    \end{figure*}

Experiments in E-TRIALS are delivered through ASSISTments. They can be assigned using learning management systems, such as Google Classroom or Canvas. As shown in Figure \ref{fig:Figure 1}, E-TRIALS supports three types of experiments: \textbf{single support}, \textbf{support comparison}, and \textbf{problem varied}. 

\textit{Single Support}: Single support studies are designed to evaluate the effects of a specific type of student support message, such as a hint or explanation. These messages can be in the form of text, images, videos, or audio. The control condition in these studies is the \textbf{best-so-far} condition. This condition represents the highest-performing student support currently in ASSISTments for that problem. 

Researchers use ASSISTments' math problem content library, which comprises problems from open-source curricula, such as Kendall Hunt, Illustrative Mathematics, and Engage NY/Eureka Math to create experimental conditions \cite{illustrative2025,pbc2025}. Once they create the experimental conditions in a problem set(s), teachers can assign the problem set(s) to their students, and in doing so, students would randomly be assigned to a condition. This process alleviates the burden of participant recruitment. 

\textit{Support Comparison}: Support comparison studies build on the single support model by allowing researchers to investigate two or more types of student support messages. Examples include comparing hints vs. explanations, hints vs. worked examples, or explanations vs. a dynamic algebraic notation system. As with single support studies, researchers can use ASSISTments' existing math content to create conditions. 
For example, \cite{smith2022impact} used a support comparison study to explore the impact of worked example extensiveness and the degree of dynamic presentation on student learning in algebra. Their study had six conditions: 1) static concise, 2) static extended, 3) sequential concise, 4) sequential extended, 5) dynamic history, and 6) dynamic no history. Data from 230 7th-12th graders revealed an overall improvement from the pretest to the posttest after viewing the worked examples, but there were no significant differences in posttest performance between worked example presentations. Meaning, the extensiveness and the degree of dynamic presentation did not impact students' posttest scores. 

\textit{Problem Varied}: Problem varied studies offer researchers the flexibility to create their own problem content and recruit participants. These studies accommodate a broader range of participants, from K-12 to higher education students. This option is suitable for researchers that want to explore research questions outside the realm of mathematics, K-12 students, or would like to use a platform other than ASSISTments to deliver the assignment. \cite{smalenberger2022college} conducted a problem varied study to investigate whether findings from middle school students can be replicated among collect level students. Both middle and college-level students were assigned to one of three tutor strategies conditions: 1) text hints, 2) text explanations, or 3) video explanations. Findings revealed that college students benefited from video explanations more than text hints. However, middle school students benefited more from video explanations than college students by 36\%. \cite{unal2020using} also used a problem varied study to design a think-aloud intervention that explored cognitive control and rule learning in a real-world setting. Their study featured probability problem sets and recruited undergraduate students as participants. 

\subsubsection{Deploying an Experiment}
E-TRIALS supports two deployment methods depending on the study type: \textbf{automatic deployment} and \textbf{self-deployment}. 
\begin{itemize}
    \item \textbf{Automatic Deployment}: Studies categorized as single support or support comparison are automatically deployed. These studies are added to ASSISTments' content library, where teachers can assign them to students.
    \item \textbf{Self-Deployment}: Problem varied studies require self-deployment because researchers need to recruit their own participation pool. Researchers collaborate with students, teachers, or schools to assign content. 
\end{itemize}
One prerequisite to deploying a study, regardless of the type, is that the researcher would need to get IRB approval from their institution. Once they attain IRB approval, then the study can be deployed and participants can begin interacting with the intervention. 

\subsubsection{Completion of Experiment}
Once a study has been deployed, researchers can request the study data after reaching a specific number of student participants or after a predetermined time frame. The choice between these options depends on the study's requirements. 

Before receiving the intervention data, researchers are required to pre-register their study at OSF.io. Once the study has pre-registration is complete, researchers receive eight distinct data files, each containing specific insights into student interactions, learning outcomes, and experimental conditions. These files provide a comprehensive view of the study's results, ranging from click stream data to problem difficulty and demographic context. 

Below is a detailed description of each data file:
\begin{enumerate}
    \item Action Logs: This file captures students' click stream data during an assignment. It records every action a student takes, such as starting, resuming, or finishing a problem, requesting a support message or submitting an answer. Each row represents a single student action. 
    \item Problem Logs: This file contains problem-level data, such as time spent on the problem, the number of attempts, and scores. ASSISTments uses two scoring metrics:
        \begin{itemize}
            \item Discrete Score: Either 0 or 1. 
            \item Continuous Score: Values can be 0, 0.33, 0.67, or 1.0. Students lose 0.33 points each time they request a hint. Meaning, the highest possible score a student can receive after viewing one hint is 0.67.
        \end{itemize}
    \item Assignment Logs: This file summarizes assignment-level data. Each row includes information on the total time spent on the assignment, the total number of attempts or hints, and the cumulative discrete score for the assignment.
    \item Prior Logs: This file provides class- and school-level data related to students' prior learning. For the class-level data, it includes the curriculum used, the class grade level, the number of students in the class, and the class's average performance. School-level data can be provided if a teacher provides their school address in their profile. Then the school address is used to retrieve school-level data from the National Center for Education Statistics (NCES) \cite{hussar2011projections}. School-level data includes demographic data, such as percentage of students receiving free or reduced lunch, ethnicity, and gender distribution, and the school's location (rural, urban, or suburban). The \textbf{school\_id} provided in this file can be used to access additional school-level, or even district-level data through NCES or the Education Data Explorer by the Urban Institute. If a teacher does not provide their school address, however, it is not feasible to link the data for their class to school-level data.
    \item Same Skill Prior Logs: This file is similar to the \textit{Prior Logs} file. However, this file focuses only on prior problems by the student that share the same skill codes as those in the experiment. 
    \item Same Skill Post Logs: This file is similar to the \textit{Same Skill Prior Logs}, but it pertains to problems completed after the experiment. 
    \item Assignment Settings: The file details the settings applied to the assignment. Some are setting a time limit or enabling the redo feature. The redo feature allows students to attempt a similar, but not the same problem again if a student fails to achieve a perfect score on the initial problem. 
    \item Redo Logs: If the redo feature is enabled, this file provides problem-level data for the redo attempts, which shows how students performed on these additional problems. 
\end{enumerate}
These eight files are available only for studies conducted with participants from ASSISTments. Studies involving external participants (e.g., college students recruited for problem-varied studies) will not include data on students' prior and post-performance in ASSISTments. 
An example of the eight data files can be viewed \href{https://osf.io/xp6ch/?view_only=c7fa2877e8224981b7d5f99dbb0a25be}{here}. 
\section{Example Studies}
\subsection{The Two Experiments}
In this section, we introduce two E-TRIALS support comparison studies that we'll use to showcase an example of analyzing E-TRIALS data to determine the efficacy of the studies. We'll explain what each experiment explored and their conditions. A link to the code is provided in the \textit{Data Cleaning \& Preparation} sub-section. Data for the two experiments can be provided upon request. 

The two experiments are summarized below:
\begin{itemize}
    \item Experiment 1 (similar vs. different example feedback): This experiment explored math learning and persistence by comparing two feedback conditions:
    \begin{enumerate}
        \item Similar Example Feedback: Feedback that aligns closely with the problem at hand. 
        \item Different Example Feedback: Feedback that diverges from the problem's context.
    \end{enumerate}      
    Students were randomly assigned to one of the two conditions. Figure \ref{fig:Figure exp_14} illustrates the differences between these feedback types. The \textbf{Similar} figure informs students in the first hint that the subsequent hint will be a new problem completely worked out, and they should reference the steps to solve the original problem. The new problem that appears in the subsequent hint is similar to the original problem, whereas in the \textbf{Different} figure, the new problem that appears in the second hint is different than the original problem. Click \href{https://osf.io/xp6ch/?view_only=c7fa2877e8224981b7d5f99dbb0a25be}{here} to view the figures in higher resolution.
    \begin{figure*}[htbp]
        \centering
        \includegraphics[width=1.0\linewidth]{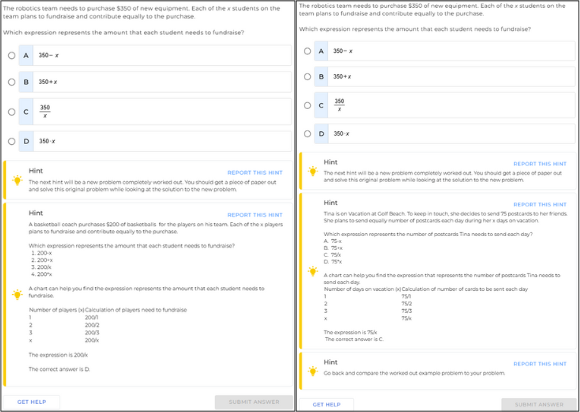}
        \caption{The support messages students received in both conditions. The left figure is the Similar condition, and the right figure is the Different condition.}
        \label{fig:Figure exp_14} 
    \end{figure*}
    \item Experiment 2 (reflect vs. regular hint): This experiment examined the effectiveness of encouraging students to reflect on their confusion before requesting an additional hint. 
         \begin{enumerate}
            \item Reflect: Students were prompted to reflect on their confusion before requesting additional hints. 
            \item Control: Students received hints as usual without any prompts to reflect. 
        \end{enumerate}  
    Students were randomly assigned to one of two conditions. Figure \ref{fig:Figure exp_16} displays the reflection prompt. The first hint in the \textbf{Reflect} figure prompts students to reflect on their knowledge, and to write down on a piece of paper the parts of the problem they understand and the parts they find confusing. In the \textbf{Control} figure, students receive hints as they normally would in ASSISTments. Click \href{https://osf.io/xp6ch/?view_only=c7fa2877e8224981b7d5f99dbb0a25be}{here} to view the figures in higher resolution. 
    \begin{figure*}[htbp]
        \centering
        \includegraphics[width=1.0\linewidth]{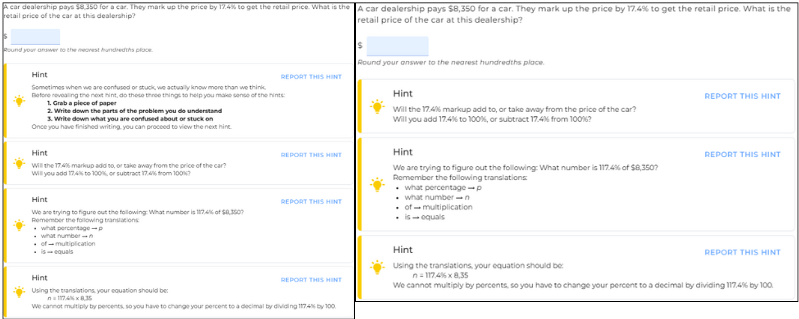}
        \caption{The support messages students received in both conditions. The left figure is the Reflect condition, and the right figure is the Control condition.}
        \label{fig:Figure exp_16} 
    \end{figure*}
\end{itemize}

\subsection{Identifying Target Populations while Avoiding Selection Bias}
Both experiments under consideration contrasted student supports---just-in-time homework help that struggling students can receive upon request. 
In both cases, our interest is in those students who requested help from the tutoring system, so we conducted our analysis on the subset of the data consisting of students who, at some point during the experiment, requested such help.

In general, depending on the design of the A/B testing platform and other specifics, this type of subsetting, as innocuous as it may seem, can lead to selection bias. For instance, say that students who request support are penalized for doing so, and the extent of the penalty is different for the two conditions in a study.
For the sake of argument, say the penalty is greater in the treatment condition than in control.
In that case, we might expect that some relatively strong students (those who are deciding between requesting support and trying to figure the problems out on their own) would be dissuaded from requesting hints if assigned to the treatment condition but not if assigned to control. If they are then dropped from the analysis sample, the average learning outcome in the treatment group will be biased downward.
This is an example of post-randomization selection bias \cite{Li03032024} that arises when researchers analyze subsets of their data that might have been affected by treatment assignment. 

Fortunately, the current version of the E-TRIALS platform has been designed to avoid this problem (though it affected earlier designs---for instance, leading \cite{patikorn2020effectiveness} to avoid subsetting). 
In order to request support, students must first press a button labeled ``Get Help." 
The appearance of this button does not depend on the type of supports that are available---in particular, it is identical for both treatment arms in our two example studies. 
Hence, before the first time during the experiment a student pressed ``Get Help," they had no way of knowing their treatment randomization. 
Therefore, the excluded subset of students who never pressed ``Get Help'' cannot be affected by treatment randomization---the subset would have included the same students under any permutation of the treatment assignment, and its exclusion does not cause selection bias.

\subsection{Data Cleaning \& Preparation}
In this section, we outline the steps taken to clean and prepare the data. The code used for this process can be accessed \href{https://osf.io/xp6ch/?view_only=c7fa2877e8224981b7d5f99dbb0a25be}{here}. 

The data cleaning process began with the \textit{Problem Logs} table. Table \ref{initial_table} shows a summary of the conditions, student and assignment scores, their means and standard deviations at the condition level, and the average number of problems assigned per condition. 

\begin{table}[h!]
\centering
\caption{Experiments Initial Summary}
\label{initial_table}
\begin{threeparttable}
 \resizebox{\textwidth}{!}{%
\begin{tabular}{l c l c l c l c l c l c l c l}
\toprule
Experiments & & Condition & & Condition Type & & Student Count & & Assignment Count & & Mean & & Std. Dev. & & Avg. Problems Assigned \\ 
\midrule
\multicolumn{1}{c}{1} & & 0         & & Similar            & & 343           & & 362              & & 0.565 & & 0.448     & & 10.06                  \\ 
            & & 1         & & Different          & & 348           & & 365              & & 0.573 & & 0.445     & & 9.81                   \\ 
            & & 2         & & Best-So-Far        & & 334           & & 362              & & 0.571 & & 0.446     & & 10.54                  \\ 
\midrule
\multicolumn{1}{c}{2} & & 0         & & Reflect            & & 517           & & 582              & & 0.702 & & 0.418     & & 10.62                  \\ 
            & & 1         & & Control            & & 488           & & 552              & & 0.708 & & 0.412     & & 10.32                  \\ 
            & & 2         & & Best-So-Far        & & 516           & & 593              & & 0.709 & & 0.412     & & 10.38                  \\ 
\midrule
\end{tabular}%
}   
\end{threeparttable}

\end{table}

To ensure we analyze the experiments correctly, several filtering steps were applied. First, we removed ASSISTments' default condition, \textit{best-so-far}. Once again, this condition represents the highest-performing student support currently in ASSISTments for that problem. Meaning, one student could receive a support message in a text format, whereas another student could receive it in a video. Due to the inconsistency, we decided to remove the condition from our analysis. 

Next, we removed students who did not request at least one hint during the assignment, as hint interaction was required to determine the effect of the experiment. This filtering step resulted in the removal of 400 students from Experiment 1 and 477 students from Experiment 2. 

Lastly, several assignments in ASSISTments contain sub-problems. A sub-problem is linked to a parent problem and builds on the same scenario or context. Sub-problems often guide students to think conceptually or to apply the result from the parent-problem to solve the sub-problem. For example, a parent problem can set out the scenario or context and then ask the student to calculate the commission from the sale. Then, the sub-problem can ask the student to use the results from the parent-problem to determine how much money the store receives after the commission is deducted. With that, it's important to note that none of the sub-problems in the two experiments were conceptual, where a student was prompted to provide a reasoning for their answer to the parent problem. Since none of the sub-problems were conceptual, we treated them as independent problems. The updated summary of the dataset after applying these filters is presented in Table \ref{filtered_table}. 

To assess student participation across the study, we calculated the attrition rate, which refers to the loss of participants over the course of a study \cite{clearinghouse2015wwc}. Attrition occurs when students who were initially randomly assigned to a condition do not provide outcome data. In our case, we measured the attrition rate using the proportion of missing values in the pre- and post-intervention continuous score variables from \textit{Same Skill Prior \& Post Logs} tables relative to the total number of observations \cite{schafer2002missing}. The attrition rates for the two experiments were low, which means the vast majority of participants completed the study and had outcome data. 

\begin{table}[h!]
\centering
\caption{Experiments Summary After Filtering \& Cleaning}
\label{filtered_table}
\begin{threeparttable}
 \resizebox{\textwidth}{!}{%
\begin{tabular}{l c l c l c l c l c l c l c l}
\toprule
Experiments & & Condition & & Condition Type & & Student Count & & Assignment Count & & Mean & & Std. Dev. & & Attrition Rate \\ 
\midrule
\multicolumn{1}{c}{1} & & 0         & & Similar           & & 157           & & 161              & & 0.427 & & 0.454     & & 0.0047         \\ 
            & & 1         & & Different         & & 134           & & 137              & & 0.396 & & 0.448     & & 0.0035         \\ 
\midrule
\multicolumn{1}{c}{2} & & 0         & & Reflect           & & 282           & & 296              & & 0.574 & & 0.458     & & 0.00004        \\ 
            & & 1         & & Control           & & 246           & & 260              & & 0.554 & & 0.455     & & 0.00005        \\ 
\midrule
\end{tabular}%
}
\end{threeparttable}
\end{table}

Next, the filtered \textit{Problem Logs} table was joined with the \textit{Assignment Logs} table to produce an assignment-level dataset. This table included details about assignment and teacher IDs, and was further aggregated to calculate the number of problems assigned to each student and the number they completed. 

\subsection{Feature Selection in Covariate Adjustment}
It is essential for students to have comparable prior knowledge and motivation to learn. The RCTs within ASSISTments use Bernoulli randomization, so students in the treatment and control arms should be comparable on average. However, in practice, slight imbalances in the students' baseline characteristics may still exist. To assess the comparability of students in the treatment and control arms of our two selected experiments, we performed a classification permutation test (CPT). This test evaluates whether the randomization procedure adequately mitigates potential confounders arising from differences in prior knowledge and learning motivation. Table \ref{tab:cov_balance} shows that while there is no strong evidence of imbalance between the treatment and control groups, some degree of covariate difference remains, suggesting that covariate adjustment is still necessary.
\begin{table}[ht]
\centering
\caption{Covariate Balance Check}
\label{tab:cov_balance}
\begin{tabular}{cc}
\toprule  Experiments & p-value \\
\midrule
1 & 0.196 \\
2 & 0.124 \\
\bottomrule
\end{tabular}
\end{table}

To minimize the influence of confounding factors, our aim is to enhance comparability through covariate adjustment. Fortunately, our data set contains a large number of covariates, which we leverage to account for variations in the students' baseline knowledge. 

First, we utilized students' prior and post-continuous scores from assignments associated with the same skill code as the experimental problems. These data were drawn from the \textit{Same Skill Prior Logs} and \textit{Same Skill Post Logs}. Second, we analyzed the \textit{Prior Logs} table to identify covariates that could be correlated with student performance. Although numerous covariates are available, feature selection is necessary to ensure computational efficiency and reduce multicollinearity, particularly for regression. However, in LOOP, which is based on Random Forest (RF) models, feature selection is handled inherently. At each split in a tree, the algorithm randomly selects a subset of features and chooses the one that best splits the data based on a predefined criterion, making feature selection a local and adaptive process. \footnote{regression and LOOP, based on Random Forest, will be discussed in detail in the next section.}

We began by selecting a set of covariates based on their variability and pairwise correlations. Specifically, we removed features with near-zero variance, as they carry little informational value, and eliminated highly correlated variables to reduce multicollinearity. Although relying solely on linear correlation does not capture nonlinear dependencies, we adopted this approach for its simplicity, ease of implementation, and compatibility with linear mixed-effects models such as \texttt{lmer}. After applying feature selection procedures, we retained a set of covariates that exhibit sufficient variability and are not highly correlated with each other. These variables were selected to ensure meaningful input into the linear mixed-effects modeling process, while reducing redundancy and potential multicollinearity. The final set includes student-level, class-level, and performance-related variables, as listed in Table~\ref{tab:final_vars}.

\begin{table}[ht]
\centering
\small
\caption{Selected Variables After Feature Selection}
\label{tab:final_vars}
\begin{tabular}{ll}
\toprule
Variable Name & Description \\
\hline
similar & Binary indicator of treatment or control condition \\
class\_xid & Unique identifier for each class \\
problems\_assigned & Number of problems assigned \\
student\_prior\_median\_first\_repsonse\_time\_task & Median first response time (prior assignments) \\
student\_prior\_median\_time\_on\_task & Median time on task (prior assignments) \\
student\_prior\_problem\_sets\_completed & Number of problem sets completed previously \\
student\_prior\_average\_attempt\_count & Average number of attempts (prior assignments) \\
student\_prior\_viewed\_assignment\_report\_percent & Percent of assignment reports viewed \\
prior\_avg\_cont\_score & Average pre-test continuous score \\
\bottomrule
\end{tabular}
\end{table}

Finally, the assignment-level dataset was combined with the prior and post-draft logs to create the final dataset used for analysis.

\section{Method}
In this section, we discuss the statistical methods employed to assess the impact of different types of tutoring support on students' proximal and distal performance. The two outcome variables that we explored are distal and proximal learning. Distal learning ($avg\_score$) refers to the knowledge that students retain after the experiment. It is measured by the average score students received on problem sets post-experiment, which assesses the same skills covered during the experiment. Proximal learning ($post\_avg\_score$) refers to the knowledge that students gain during the experiment. It is measured by the average continuous score that the students received during the set of experiment problems. We also provide the rationale for selecting these methods, emphasizing their respective advantages and limitations. The methods are presented in formal mathematical terms for clarity and rigor.

Consider an experiment with $N$ participants, indexed as $i = 1,2,...,N$. Each participant can be assigned to either treatment or control arm at random, and we denote $T_{i}$ following Bernoulli distribution be the $i$ th participant's treatment assignment, that is,
\begin{align}
    T_{i} &\perp T_{j}, \quad i \neq j \\ \notag
    P(T_{i} = t) &= p_{i}^t (1-p_{i})^{1-t}, \quad t \in \{0, 1\}
\end{align}
where $T_{i} = 1$ when the $i$ th participant is assigned to treatment arm and $T_{i} = 0$ when control arm. Furthermore, we observe outcomes $Y_{i}$ and $q$-dimensional baseline covariates $Z_{i}$ besides treatment assignment $T_{i}$ for each participant.

Assume each participant has two nonrandom potential outcomes-- treatment potential outcomes $t_{i}$ and control potential outcomes $c_{i}$ \cite{rubin1974estimating,b2eb7cdb-e49f-3649-82c7-5c6d5ed5a61e}. We can write observed outcomes $Y_{i}$ as
\begin{equation}
    Y_{i} = T_{i}t_{i} + (1-T_{i})c_{i}
\end{equation}
Individual Treatment Effect (ITE) $\tau_{i}$ can be written as
\begin{equation}
    \tau_{i} = t_{i} - c_{i}
\end{equation}
and the Average Treatment Effect (ATE) $\bar{\tau}$ as
\begin{equation}
    \bar{\tau} = \frac{1}{N}\sum_{i=1}^{N}\tau_{i}
\end{equation}
which is the main parameter we're interested in estimating.

Below we introduce LOOP estimator $\hat{\tau}_{LOOP}$ along with the other two common ATE estimators: Student's t-test estimator $\hat{\tau}_{t-test}$ and Regression estimator $\hat{\tau}_{Reg}$. We also explore their strengths and limitations.

\subsection{Student's t-test}

$\hat{\tau}_{t-test}$ represents the difference between the average of the observed treatment outcomes and the average of the observed control outcomes.
\begin{equation}
    \hat{\tau}_{t-test} = \frac{1}{n}\sum_{i=1}^{n}Y_{i} - \frac{1}{N-n}\sum_{i=1}^{N-n}Y_{i}
\end{equation}
where $n$ is the number of participants in the treatment arm. It provides an unbiased estimate of the average treatment effect with $0<n<N$.

\subsection{Regression}

Outcomes $Y_{i}$ and $q$-dimensional baseline covariates $X_{i}$ are available in E-Trials, enabling the implementation of covariate adjustments for estimating the ATE. Regression estimators $\hat{\tau}_{Reg}$ naturally arise when considering covariate adjustment.
\begin{equation}
    Y_{ij} = \tau_{\text{Reg}} \cdot T_{ij} + \mathbf{X}_{ij}' \boldsymbol{\beta} + u_j + \epsilon_{ij}
\end{equation}
where \( Y_{ij} \) is the outcome for student \( i \) in class \( j \), \( T_{ij} \) is a binary treatment indicator, \( \mathbf{X}_{ij} \) is a vector of observed covariates, \( u_j \sim \mathcal{N}(0, \sigma_u^2) \) is a class-level random intercept, and \( \epsilon_{ij} \sim \mathcal{N}(0, \sigma^2) \) is the individual error term. Under the assumptions of no unmeasured confounding given \( \mathbf{X}_{ij} \), correct model specification, and no interference, the coefficient \( \hat{\tau}_{\text{Reg}} \) provides an unbiased estimate of ATE, \( \tau \).

By adjusting for treatment \( T_{ij} \) and covariates \( \mathbf{X}_{ij} \), \( \hat{\tau}_{\text{Reg}} \) typically improves precision over the unadjusted estimator \( \hat{\tau}_{\text{t-test}} \). However, it may be biased if \( \mathbf{X}_{ij} \not\perp T_{ij} \) due to imperfect randomization, and it assumes linearity between covariates and outcomes, potentially overlooking nonlinear effects.

\subsection{LOOP}

To address the limitations of the Regression estimator, we propose using LOOP to estimate $\tau$. $\hat{\tau}_{LOOP}$ offers an unbiased estimate of ATE while incorporating covariate adjustments \cite{wu2018loop}. We present the steps of the LOOP methodology. A key quantity, $m_{i}$, is defined for each participant as follows:
\begin{equation}
    m_{i} = (1-p_{i})t_{i} + p_{i}c_{i}
\end{equation}
and a (signed) inverse probability weight $U_{i}$, defined as follows:
\begin{equation}
U_i =
\begin{cases}
\frac{1}{p_i}, &  T_i = 1, \\
-\frac{1}{1-p_i}, &  T_i = 0.
\end{cases}
\end{equation}
and note that $U_{i}$ has expectation $0$.

Also we have
\begin{align}
    Y_iU_i =
\begin{cases}
\frac{t_i}{p_i}, &  T_i = 1, \\
-\frac{c_i}{1-p_i}, &  T_i = 0.
\end{cases}
\end{align}

Now, consider $\hat{\tau}_{i} = (Y_{i} - \hat{m}_{i})U_{i}$ as an unbiased estimator of ITE.

When $\hat{m}_{i} \perp U_{i}$, $\hat{\tau}_{i}$ can serve as an unbiased estimator of ITE. This can be shown as follows:
\begin{align}
    E(\hat{\tau}_{i}) &= E[(Y_{i}-\hat{m}_{i})U_{i}] \notag \\
    &= E(Y_{i}U_{i}) - E(\hat{m}_{i})E(U_{i}) \notag \\
    &= \frac{t_i}{p_i}P(T_i = 1) + \frac{-c_i}{1-p_i}P(T_i = 0) + E(\hat{m_i})\cdot 0 \notag \\
    &= t_i - c_i + 0 \notag \\
    &=\tau_{i}
    \label{eq:unbias}
\end{align}
Additionally, we can demonstrate that $\hat{\tau}_{i}$ has reduced variance compared to $\hat{\tau}_{\text{t-test}}$, provided that $\hat{m}_{i}$ is a good approximation of $m_i$, which is satified when $\hat{t}_{i}$ and $\hat{c}_i$ closely approximates $t_i$ and $c_i$. The variances can be calculated as follows:
\begin{equation}
    Var(\hat{\tau}_{i}) =\frac{1}{p_{i}(1-p_{i})}E[(\hat{m}_{i}-m_{i})^{2}]
\end{equation}
To summarize, $\hat{\tau}_{i}$ is not only unbiased but also exhibits reduced variance when (a) $\hat{m}_{i} \perp T_{i}$ and (b) $\hat{m}_{i}$ serves as an accurate estimator of $m_{i}$.

We now introduce Leave-One-Out imputation. Let the LOOP estimator of ATE, $\hat{\tau}_{LOOP}$, be denoted as:
\begin{equation}
    \hat{\tau}_{LOOP} = \frac{1}{N}\sum_{i=1}^{N}\hat{\tau}_{i}
\end{equation}
where $\hat{\tau}_{i} = (Y_{i}-\hat{m}_{i})U_{i}$, and $\hat{m}_{i}$ is obtained using Leave-One-Out imputation, i.e., we sequentially exclude each observation $i$ and use the remaining $N-1$ observations to impute $t_{i}$ and $c_{i}$. The imputation method is flexible and can be customized as needed. We define $\hat{m}_{i}$ as follows:
\begin{equation}
    \hat{m}_{i} = (1-p_{i})\hat{t}_{i} + p_{i}\hat{c}_{i}
\end{equation}
By leaving out the $i$-th observation when imputing $\hat{m}_{i}$, we ensure that $\hat{m}_{i} \perp T_{i}$, guaranteeing that $\hat{\tau}_{i}$ remains unbiased. Consequently, $\hat{\tau}_{LOOP}$ is also unbiased, regardless of the algorithm used to impute $t_{i}$ and $c_{i}$. Note that both $\hat{t}_{i}$ and $\hat{c}_{i}$ are imputed, even when one is observed, to maintain the independence $\hat{m}_{i} \perp T_{i}$.

When it comes to methods for imputing potential outcomes, simply ignoring covariates and taking the mean of the observed outcomes in each treatment group makes the LOOP estimator equivalent to the student's t-test estimator. However, the LOOP estimator outperforms $\hat{\tau}_{\text{t-test}}$ as long as the imputation of potential outcomes improves beyond the baseline mean imputation approach. In the work by Wu and Gagnon-Bartsch \cite{wu2018loop}, Random Forest (RF) is recommended for imputing $t_{i}$ and $c_{i}$. This is because RF, as an enhancement over individual decision trees, not only improves accuracy, but also provides automatic variable selection, resulting in more computational efficiency.

In this paper, we calculate the t-test estimator $\hat{\tau}_{\text{t-test}}$, the Regression estimator $\hat{\tau}_{\text{Reg}}$, and the LOOP estimator $\hat{\tau}_{\text{LOOP}}$ to demonstrate that the LOOP estimator is an unbiased estimator of the ATE and achieves greater precision by adjusting for baseline covariates shown in equation \ref{eq:unbias}. Furthermore, when using the LOOP estimator with RF, we can identify non-parametric relationships within the data that the Regression estimator fails to capture.

\section{Results \& Discussions}
In this section, we present the results of the examples of experiments. Tables \ref{results_14} and \ref{result_16} summarize the ATE estimates, $\hat{\tau}_{\text{t-test}}$, $\hat{\tau}_{\text{Reg}}$, and $\hat{\tau}_{\text{LOOP}}$, together with their respective variations and significance measures.

\begin{figure}[ht]
    \centering
    \includegraphics[width=\textwidth]{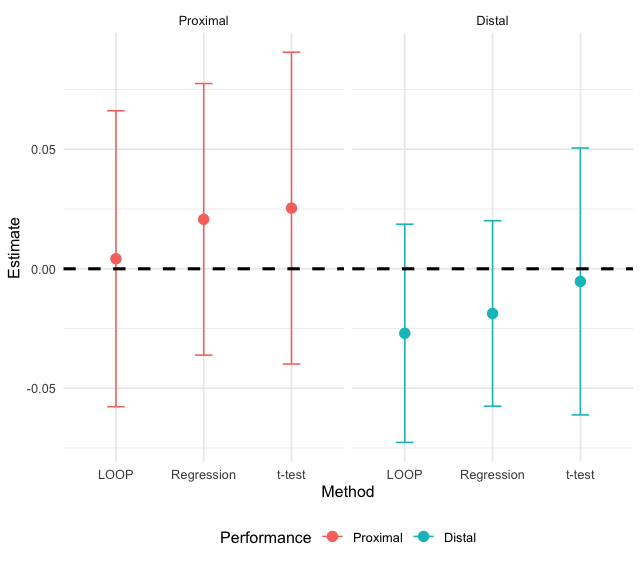} 
    \caption{Experiment 1: Confidence Intervals}
    \label{fig:CI14}
\end{figure}

First, we present the results of the analysis for Experiment 1, shown in Table \ref{results_14}, which aims to determine whether providing students with feedback that aligns closely with the problem at hand or diverges from its context influences their proximal and distal performance. We include estimates, standard errors, and $p-value$ of the results, accompanied by a Figure \ref{fig:CI14} visualizing the confidence intervals. This visualization helps us explore whether there is a likely significant positive or negative effect.

\begin{table}[ht]
\centering
\caption{Experiment 1 \& 2: Comparison of Treatment Effect Estimation}
\scriptsize
\begin{subtable}[t]{0.5\textwidth}
\centering
\caption{Experiment 1: Similar vs. Different Feedback}
\label{results_14}
\begin{tabular}{llccc}
\toprule
Perf. & Method & Est. & SE & p-value \\
\midrule
\multirow{3}{*}{Prox.} 
  & $\hat{\tau}_{t\text{-test}}$ & 0.0254 & 0.0333 & 0.4461 \\
  & $\hat{\tau}_{\text{Reg}}$    & 0.0207 & \textbf{0.0290} & 0.4757 \\
  & $\hat{\tau}_{\text{LOOP}}$   & 0.0042 & 0.0316 & 0.8957 \\
\midrule
\multirow{3}{*}{Distal} 
  & $\hat{\tau}_{t\text{-test}}$ & -0.0053 & 0.0285 & 0.8516 \\
  & $\hat{\tau}_{\text{Reg}}$    & -0.0187 & \textbf{0.0198} & 0.3443 \\
  & $\hat{\tau}_{\text{LOOP}}$   & -0.0270 & 0.0233 & 0.2474 \\
\bottomrule
\end{tabular}
\vspace{1mm}
\\
\textit{Smallest SE per group is in bold.}
\end{subtable}
\hfill
\begin{subtable}[t]{0.45\textwidth}
\centering
\caption{Experiment 2: Reflect vs. Regular Hint}
\label{result_16}
\begin{tabular}{llccc}
\toprule
Perf. & Method & Est. & SE & p-value \\
\midrule
\multirow{3}{*}{Prox.} 
  & $\hat{\tau}_{t\text{-test}}$ & 0.0055 & 0.0248 & 0.8237 \\
  & $\hat{\tau}_{\text{Reg}}$    & 0.0084 & 0.0174 & 0.6297 \\
  & $\hat{\tau}_{\text{LOOP}}$   & -0.0196 & \textbf{0.0173} & 0.2594 \\
\midrule
\multirow{3}{*}{Distal} 
  & $\hat{\tau}_{t\text{-test}}$ & -0.0060 & 0.0191 & 0.7555 \\
  & $\hat{\tau}_{\text{Reg}}$    & -0.0073 & \textbf{0.0108} & 0.5031 \\
  & $\hat{\tau}_{\text{LOOP}}$   & -0.0136 & 0.0115 & 0.2373 \\
\bottomrule
\end{tabular}
\vspace{1mm}
\\\textit{Smallest SE per group is in bold.}
\end{subtable}
\end{table}

From Table \ref{results_14}, we observe no significant difference between the similar and different feedback groups. However, after covariate adjustment, the standard errors, as indicated by the Std.Error column, are reduced. The results show that $\hat{\tau}_{\text{Reg}}$ achieves the smallest standard error for both proximal and distal performance. However, it is important to note that $\hat{\tau}_{\text{Reg}}$ might be biased, while $\hat{\tau}_{\text{LOOP}}$ is unbiased and exhibits the second smallest standard error. These findings confirm that while both Regression and LOOP provide more precise estimators compared to the student's t-test, only LOOP offers an unbiased estimate based on equation \ref{eq:unbias}. Thus, it is reasonable to select LOOP as the preferred estimator to estimate ATE.
From Figure \ref{fig:CI14}, we observe that all confidence intervals include $y=0$, indicating that none of the estimates are statistically significant. However, there is a noticeable trend: for proximal performance, students receiving different feedback tend to perform better, whereas for distal performance, they tend to perform worse. This is reflected in the proportion of the spread of the confidence intervals, which leans positive for proximal performance and negative for distal performance. Additionally, we can see that the lengths of the confidence intervals for $\hat{\tau}_{Reg}$ and $\hat{\tau}_{LOOP}$ are shorter than those for the $\hat{\tau}_{t-test}$, suggesting that regression and LOOP provide more precise estimates compared to the student's t-test.

\begin{figure}[ht]
    \centering
    \includegraphics[width=\textwidth]{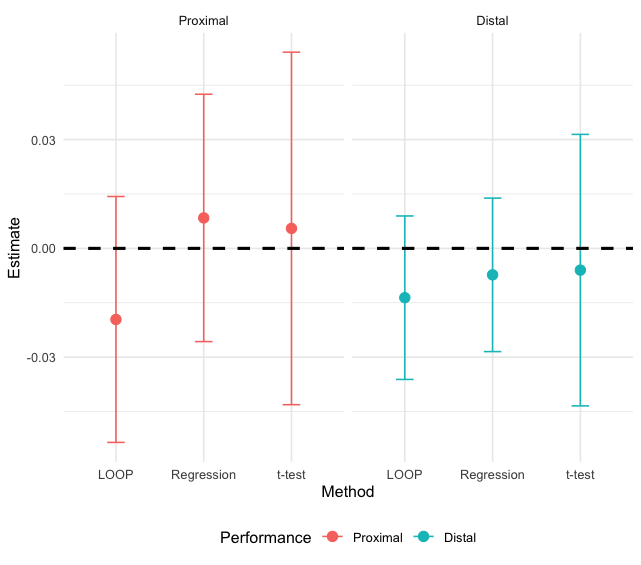} 
    \caption{Experiment 2: Confidence Intervals}
    \label{fig:CI16}
\end{figure}

Next, we examine the insights gained from Experiment 2, which investigated whether prompting students to reflect on their confusion before accessing a hint has an impact on students' performance. The findings are summarized in Table \ref{result_16} illustrated in Figure \ref{fig:CI16}.
Similar to the findings from the previous experiment, $\hat{\tau}_{LOOP}$ is the recommended estimator due to its combination of a small standard error and unbiased property. Additionally, Figure \ref{fig:exp1416_comparison} right hand suggests that prompting students to reflect is more likely to enhance their performance in the proximal period, but may negatively affect their performance in the distal term—although this effect is not statistically significant at the 5\% level.

By reviewing these two experiments, we observe the clear advantages of using $\hat{\tau}_{LOOP}$ to estimate ATE. This estimator is both precise and easy to apply, making it a versatile choice for other experiments involving comparative tutoring designs. One notable observation is that the distal performance estimates have smaller standard errors and narrower confidence intervals. At first glance, this may seem counterintuitive, given that distal performance involves long-term prediction, which is typically more challenging. However, this phenomenon becomes understandable when considering that proximal performance—a variable highly correlated with distal outcomes—was included as a super covariate in the distal performance model. Since students’ current performance strongly predicts their future performance, incorporating this information enhances the precision of the distal performance estimates.

Furthermore, to gain a deeper understanding of why these interventions enhance or impair students' performance during the proximal or distal periods, additional analyses are needed. These could involve controlling for additional confounding factors or conducting psychological analyses to uncover the underlying mechanisms. Nonetheless, LOOP can be widely applied to various support comparison scenarios and provides valuable insights that can inform the design of future educational experiments.


\section{Limitations, Future Plans, \& Conclusions}
\subsection{Limitations}
In this paper, we utilized data from E-TRIALS to evaluate whether different tutoring strategies have a significant positive or negative impact on students' proximal and distal performance, where the LOOP estimator is recommended for estimating the ATE due to its reduced standard error and unbiasedness. However, there are certain limitations associated with the methods employed.
\begin{enumerate}
    \item Although the vanilla ATE estimation code works well for support-comparison RCTs with student-level randomization, modifications are required when randomization is applied at a higher level (e.g., class or school). Additional adjustments may also be necessary for single-support studies, problem-varied studies, or experiments conducted in a different CBLP. However, as long as the provided data follow the E-TRIALS format, these methods can still be applied effectively.
    \item For the LOOP estimator to achieve low variance, it is essential that \( \hat{m}_i \) is a reliable estimate of \( m_i \), which requires accurate estimation of both \( c_i \) and \( t_i \). Additionally, to ensure a variance reduction relative to \( \hat{\tau}_{t-test} \), it is necessary that \( m_i \perp T_i \). Satisfying both conditions can sometimes be challenging.
    \item The LOOP method cannot handle categorical variables with more than $53$ levels, as observed when including $class\_xid$ as a covariate to adjust for the baseline performance of the students in Experiment 2. Further revisions to the LOOP package are necessary to overcome this limitation.
\end{enumerate}

\subsection{Future Plans}
The method and code discussed in this paper are designed to support comparison studies. One future direction is to develop a similar framework for single support studies. However, it is not feasible to create a generalized framework for problem-varied studies due to the flexibility these studies provide researchers in designing and conducting experiments. This autonomy makes it challenging to standardize the process, as each study may require unique adaptations. 

Another future plan is to build a Shiny application to fully automate the data cleaning and modeling process. This application would allow users to upload their CSV files and select from a predefined list of outcome variables. By automating this process, the Shiny application could reduce the time and effort required to evaluate experiments. 

\subsection{Conclusion}
Despite the significant increase in the use of computer-based learning platforms, researchers still lack adequate tools to address the diverse challenges students face in these environments. These challenges range from a one-size-fits-all instructional approach to the delivery of culturally insensitive feedback messages. Such issues are a testimony to the urgent need for tools that can enable researchers to identify and implement effective solutions. 

In this paper, we introduced E-TRIALS, a tool designed to help researchers make data-driven decisions to improve student performance in CBLPs. E-TRIALS supports multiple study types, each with its own advantages and disadvantages, enabling researchers to conduct experiments tailored to their specific needs. Our work demonstrates the potential of E-TRIALS to advance research in the realm of learning sciences and support the design of evidence-based CBLPs. 
\section{Acknowledgments}
We would like to thank NSF (e.g., 2118725, 2118904, 1950683, 1917808, 1931523, 1940236, 1917713, 1903304, 1822830, 1759229, 1724889, 1636782, \& 1535428), IES (e.g., R305N210049, R305D210031, R305A170137, R305A170243, R305A180401, \& R305A120125), GAANN (e.g., P200A180088 \& P200A150306), EIR (U411B190024 \& S411B210024), ONR (N00014-18-1-2768), NHI (R44GM146483), and Schmidt Futures. None of the opinions expressed here are that of the funders.


\bibliographystyle{acmtrans}
\bibliography{./ref}

\end{document}